\documentclass[twocolumn,prl,aps,showpacs]{revtex4}
\usepackage{graphicx}

\begin{document}

\title {Exact Non-Stationary Probabilities in the
Asymmetric Exclusion Process on a Ring}

\author{V.B. Priezzhev$^{1,2,*}$}

\affiliation{$^1$Max-Planck Institute for the Physics of Complex Systems, N\"othnitzer strasse 38, D 01187 Dresden, Germany}

\affiliation{$^2$Bogolubov Laboratory of Theoretical Physics,
Joint Institute of Nuclear Research, 141980 Dubna, Russia }


\begin{abstract}
~\newline~\newline
By a geometrical treatment of the Bethe ansatz, we obtain an exact solution
for the totally asymmetric exclusion process on a ring. We derive an explicit
determinant expression for the non-stationary conditional probability
$Prob(x_1,...,x_P;t|x_1^0,...,x_P^0;0)$ of finding $P$  particles on sites
$x_1,...,x_P$ at time $t$ provided they are on sites  $x_1^0,...,x_P^0$
at time $t=0$.
\end{abstract}

\maketitle

The one-dimensional asymmetric exclusion process (ASEP) has been intensively
studied as one of the simplest examples of a system with stochastic dynamics
and exclusion interaction \cite{Liggett},\cite{Spohn},\cite{Krug}.
In the steady state, many properties
of the ASEP have been calculated exactly (see,e.g.,
\cite{Gva},\cite{DerEvMuk},\cite{JanLeb},\cite{Derrida} and references
therein). The structure of states out of stationarity is more complicated and
description of dynamics is harder to obtain \cite{Dhar},\cite{DerMal}. In
general, evaluation of the conditional probability of finding
$P$ particles on lattice sites $x_1,...,x_P$ at time $t$ with initial
occupation $x_1^0,...,x_P^0$ at time $t=0$, needs solving a non-trivial master
equation. The solution of the master equation for the ASEP with a finite
number of particles  on an infinite lattice was obtained in \cite{Schutz}
where an explicit determinant expression for the probability
$Prob(x_1,...,x_P;t|x_1^0,...,x_P^0;0)$ was derived. At the same time,
a non-stationary solution for the finite density of particles is desirable
for considering the evolution of a system from arbitrary initial conditions to
the steady state.

It is the aim of this Letter to present an exact expression for the
conditional probability of finding $P$ particles on an finite ring of $L$
sites after the evolution with the ASEP dynamics during finite time $t$.
In the theory of exactly soluble models, the determinant solutions on finite
two-dimensional domains permitting a dynamical interpretation in space-time
coordinates are known either for so-called free-fermion models
\cite{Baxter}, or for particular domain-wall boundary conditions
\cite{Korepin},\cite{Izergin}. The ASEP does not belong
to the free-fermion class and cannot be restricted to a domain wall geometry.
However, we will see that a specific property of the pair interaction  in the
ASEP allows one to consider the model on the finite ring and
express the solution in a form of determinant  of a $P\times P$ matrix.

Instead of explicitly solving the master equation on the ring, we use here
a geometrical treatment of the Bethe ansatz which allows us to analyse a
tangled system of allowed and forbidden trajectories of particles.
To make the presentation more transparent, we consider a discrete space-time
version of the totally ASEP. The continuous time limit then follows from
the final expressions by the straightforward substitution of the Bernoulli
distribution by its Poisson analogue.

Consider the  triangle lattice $\Lambda$ obtained from the square
lattice by adding a diagonal between the upper left corner and the lower right
corner of each elementary square.The lattice is periodic in the
horizontal direction with the period $L$. Let $(x,t)$ be integer coordinates of
a particle on $\Lambda$ where the vertical time axis is directed down and the
horizontal space axis is directed right. A trajectory of the particle is a
sequence of connected vertical and diagonal bonds of $\Lambda$.
Each diagonal bond corresponds to one jump of the particle to its right for a
unit of time and has a statistical weight $z$. The vertical bond corresponds
to a stay at given  site for the unit time interval and has a statistical
weight $y$. The generating function of all possible free trajectories of one
particle for time $t$ is $(z+y)^t$. For trajectories starting at an initial
position $(x^0,0)$ and ending at  $(x,t)$, we define

\begin{equation}
B(N,t)=
\left(
\begin{array}{c}
t\\
N
\end{array}
\right)
z^Ny^{t-N}
\label{1}
\end{equation}
where $N=x-x^0$ is the travelled distance.
To provide $B(N,t)$ with probabilistic meaning, we put $z<1$, $z+y=1$.
Then, $B(N,t)$ is the probability to reach  $(x,t)$ from $(x^0,0)$
for $t$ time steps provided $z$ is the probability of one step in the right
direction. In the continuum limit, $z \rightarrow 0$, $t  \rightarrow \infty$,
we have
\begin{equation}
B(N,t)=\frac{e^{-t}t^N}{N!}
\label{2}
\end{equation}
where $t$ is rescaled continuous time $tz \rightarrow t$. For $N<0$, we
put $B(N,t)=0$.

The ASEP in discrete space-time can be defined as follows. Consider
trajectories of $P$ particles on the lattice $\Lambda$ starting at points
$(x_1^0,0),...,(x_P^0,0)$, $0 \leq x_1^0 < x_2^0 <...<x_P^0 < L $ and ending
at points $(x_1,t),...,(x_P,t)$, $0 \leq x_1 < x_2 <...<x_P < L $ after an
arbitrary number of rotations around the ring.
The exclusion rules read: (a) trajectories of particles do not intersect;
(b) for an arbitrary elementary square of $\Lambda$, if two vertical bonds are
occupied by adjacent trajectories, the weight of the left bond is changed from
$y$ to 1. The rule (a) is the usual condition of occupation of every site by
at most one particle. The rule (b) implies that the moving particle
stays at a given site with probability 1 if the target site is occupied by a
standing particle.

The generating function
$G_P(x_1,...,x_P;t|x_1^0,...,x_P^0;0)$  of the discrete ASEP is the sum over
all trajectories of $P$ particles allowed by the exclusion rule (a)
and weighted according to the rule (b). Due to the exclusion rules, $G_P$
is $Prob(x_1,...,x_P;t|x_1^0,...,x_P^0;0)$.

To formulate the main result of this Letter, consider the function
\cite{Schutz}

\begin{equation}
F_m(x_i^0,x_j|t)=\sum_{k=0}^{\infty}\left(
\begin{array}{c}
k+m-1\\
m-1
\end{array}
\right)F_0(x_i^0-k,x_j|t)
\label{3}
\end{equation}
if integer $m>0$, and
\begin{equation}
F_m(x_i^0,x_j|t)=\sum_{k=0}^{-m}(-1)^k\left(
\begin{array}{c}
-m\\
k
\end{array}
\right)F_0(x_i^0-k,x_j|t)
\label{4}
\end{equation}
if integer $m<0$.
For $m=0$,
\begin{equation}
F_0(x_i^0,x_j|t)=B(x_j-x_i^0,t)
\label{5}
\end{equation}
where $B(N,t)$ is given by Eq.(\ref{1}).
Then, for integer $L>1$ and $0<P<L$ the generating function
$G_P(x_1,...,x_P;t|x_1^0,...,x_P^0;0)$
is
\begin{equation}
G_P=\sum_{n_1=-\infty}^{\infty}...\sum_{n_P=-\infty}^{\infty}
(-1)^{\sum_{i<j}|n_i-n_j|} \det {\bf M}
\label{6}
\end{equation}
Elements of the $P\times P$ matrix {\bf M} are
\begin{equation}
M_{ij}=F_{s_{ij}}(x_i^0,x_j+n_jL|t)
\label{7}
\end{equation}
where
\begin{equation}
s_{ij}=(P-1)n_j-\sum_{k\neq j}n_k+j-i
\label{8}
\end{equation}
In the continuous limit, $B(N,t)$ is given by Eq.(\ref{2}) and $G_P$ coincides
with the probability $Prob(x_1,...,x_P;t|x_1^0,...,x_P^0;0)$ of the totally
ASEP in its standard formulation.

The derivation of this result is based on a common property of integrable
models admitting of a two-dimensional graphic representation: interchanging of
end points of two trajectories leads to their crossing. The idea of the
Bethe-ansatz is to represent trajectories of interacting particles by a set of
free trajectories given by Eq.(\ref{1}) or Eq.(\ref{2}). Then, using the
one-to one correspondence between intersections and permutations, one can
reduce enumerating  all interacting trajectories to a proper choice of signs of
permutations.

We start with the case of two particles $P=2$. According to the Bethe ansatz,
we try to represent the motion of particles by free trajectories from
$(x_i^0,0)$ to $(x_i,t), i=1,2$. Consider an elementary square of $\Lambda$
with space coordinates $x$ of the left side and $x+1$ of the right side.
Assume that particles come for the first time to neighboring sites at a moment
$t^{'}$ when one trajectory reaches the site $(x,t^{'})$ from $(x_1^0,0)$ and
another reaches the site $(x+1,t^{'})$ from $(x_2^0,0)$. To ensure  correct
weights of the next steps of interacting particles after moment $t^{'}$,
we have to exclude two possibilities from all continuations of trajectories
(Fig.1a):

(i)for the first particle,the step from $(x,t^{'})$ to $(x+1,t^{'}+1)$ with
weight $z$ and then from $(x+1,t^{'}+1)$ to $(x_1,t)$; for the second
particle, the step from $(x+1,t^{'})$ to $(x+1,t^{'}+1)$ with weight $y$ and
then from $(x+1,t^{'}+1)$ to $(x_2,t)$.

(ii)for the first particle,the step from $(x,t^{'})$ to $(x,t^{'}+1)$ with
weight $y-1=-z$ and then from $(x,t^{'}+1)$ to $(x_1,t)$; for the second
particle, the step from $(x+1,t^{'})$ to $(x+1,t^{'}+1)$ with weight $y$ and
then from $(x+1,t^{'}+1)$ to $(x_2,t)$.

\begin{figure}[h]
\includegraphics[width=76mm]{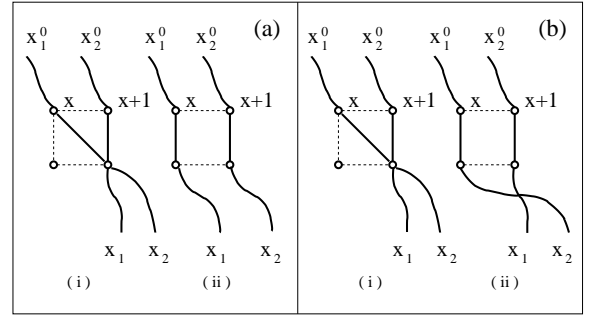}
\caption{\label{fig1}The interaction between two trajectories (see text).}
\end{figure}

Case (i) is the forbidden step of the first particle toward the standing
second particle. Case (ii) is a correction of the weight  of the vertical step
of the first particle which must be 1 instead of $y$ according to the ASEP
rule (b). The generating function of paths of the first particle in case (i)
is a product of three factors $B(x-x_1^0,t^{'}) z B(x_1-x-1,t-t^{'}-1)$.
Using symbolic notations $W(a,x|z|x+1,b)$  for the
generating function of trajectories passing points  $a,x,x+1,b$  at
moments $0,t^{'},t^{'}+1,t$ and making a step with weight $z$ between
$t^{'}$ and $t^{'}+1$, we can write the
contribution from  diagram (i) in the form
\begin{equation}
W_1=W(x_1^0,x|z|x+1,x_1)W(x_2^0,x+1|y|x+1,x_2)
\label{9}
\end{equation}
The contribution from the diagram (ii) is
\begin{equation}
W_2=-W(x_1^0,x|z|x,x_1)W(x_2^0,x+1|y|x+1,x_2)
\label{10}
\end{equation}

Consider now the trajectories where the end points are interchanged (Fig 1b).
The contribution from these diagrams is
$W(x_1^0,x|z|x+1,x_2)W(x_2^0,x+1|y|x+1,x_1)
-W(x_1^0,x|z|x,x_2)W(x_2^0,x+1|y|x+1,x_1)$.
We are going to take the diagrams in Fig.1b with opposite signs to cancel
$W_1+W_2$.
The left diagrams in Fig.1a and Fig.1b are equivalent, however the right ones
are different. To cancel all unwanted diagrams, we add to
the diagrams in Fig.1b a set of auxiliary trajectories. Namely, add to
trajectories of the second particle those starting in point $x_2^0-1$ and
taken with minus sign. Also, we add to trajectories of the first particle
a set of trajectories starting  at the points shifted along the ring in
negative directions: $x_1^0-1$, $x_1^0-2$,
$x_1^0-3$,... . If a shift exceeds $kL$, where $k>0$ is integer, the trajectory
wraps the
cylinder $\Lambda$ $k$ times.Then,the  contribution from diagram (i) in Fig.1b
will be
\begin{equation}
\tilde{W}_1=W_1^{+}W_1^{-}
\label{11}
\end{equation}
where
\begin{equation}
W_1^{+}=\sum_{k=0}^{\infty}W(x_1^0-k,x-k|z|x+1-k,x_2)
\label{12}
\end{equation}
and
\begin{equation}
W_1^{-}=W(x_2^0,x+1|y|x+1,x_1)-W(x_2^0-1,x|y|x,x_1)
\label{13}
\end{equation}
Correspondingly, for the diagram (ii) in Fig.1b, we have
\begin{equation}
\tilde{W}_2=W_2^{+}W_2^{-}
\label{14}
\end{equation}
where
\begin{equation}
W_2^{+}=-\sum_{k=0}^{\infty}W(x_1^0-k,x-k|z|x-k,x_2)
\label{15}
\end{equation}
and $W_2^{-}=W_1^{-}$.
Taking into account that generating functions of trajectories
from $(x_i^0-k,0)$, $i=1,2$, to $(x-k,t^{'})$ are equal for all $k$
due to translation invariance, one can check the identity
\begin{equation}
W_1+W_2-\tilde{W}_1-\tilde{W}_2=0
\label{16}
\end{equation}
comparing all positive and negative terms.

Consider first the evaluation of generating function
$G_2(x_1,x_2;t|x_1^0,x_2^0;0)$ in the case $L \gg t$ and $L \gg x_2^0>x_1^0
\gg 0$ which is equivalent to the ASEP on an infinite lattice solved in
\cite{Schutz}.
In this case,
\begin{equation}
\begin{array}{ll}
G_2=B(x_1-x_1^0,t)B(x_2-x_2^0,t)&-(B(x_1-x_2^0,t)\\
-B(x_1-x_2^0+1,t))&\hspace{-2cm}\sum_{k=0}^{\infty}B(x_2-x_1^0+k,t)
\label{17}
\end{array}
\end{equation}
Indeed, the first term in Eq.(\ref{17}) generates all possible free
trajectories  from initial to end points. When one particle approaches
another, the second term produces trajectories cancelling unwanted terms. On
the other hand, the order of starting and ending points in the second term is
interchanged. Therefore, each trajectory from the second term starting at
$x_1^0-k$ or $x_2^0-k$ approaches at least once the point $(x-k,t^{'})$ or
$(x+1-k,t^{'})$ where it participates in the cancellation procedure.

Each free trajectory from $a$ to $b$ making the vertical step at the collision
site $x$ can be decomposed into two parts $W(a,x|1|x,b)+W(a,x|-z|x,b)$. The
second part is unwanted and is cancelled, the first one corresponds to
trajectories which continue with true weights up to the next collision.
As the second term in Eq.(\ref{17}) contains  intersecting trajectories only,
all of them will be cancelled eventually and only true allowed trajectories
from the first term survive.

The ASEP on the ring has several peculiarities. To fix them, let us note that
two intersecting trajectories are non-equivalent: one of them belongs to the
overtaking particle and we may call it "active". On the contrary, the second
particle can be called "passive". In the case of infinite lattice, the active
and passive trajectories are ordered: for each pair $i,i+1$, the trajectory
of $i$-th particle with respect to $i+1$ particle is always active.
On the ring, each of two trajectories can be active or passive independently
on initial conditions. Moreover, one trajectory can intersect another $m$
times if the numbers of rotations differ by $m$ for two  particles.

Assume, that the trajectory of given particle has $m$ active intersections. It
means that it participates $m$ times in the cancellation procedure and its
starting point is shifted $m$ times to arbitrary distances in the negative
direction of the ring.
As a result, the auxiliary set associated with the
free trajectory between points $x_i^0$ and $x_j$ becomes
\begin{equation}
B(x_j-x_i^0,t) \rightarrow \sum_{k=0}^{\infty}\left(
\begin{array}{c}
k+m-1\\
m-1
\end{array}
\right)B(x_j-x_i^0+k,t)
\label{18}
\end{equation}
because the shift by $k$ positions for $m$ attempts can be done in
$(k+m-1)!/(m-1)!k!$ ways. The above can be expressed in an operator form
\begin{equation}
B(x_j-x_i^0,t) \rightarrow \frac{1}{(1-\hat{a}_i)^m}B(x_j-x_i^0,t)
\label{19}
\end{equation}
where the operator $\hat{a}_i$ shifts $x_i^0$ by one step in the negative
direction.
Similarly,for trajectories having $m$ passive
intersections we get
\begin{equation}
B(x_j-x_i^0,t) \rightarrow \sum_{k=0}^{m}(-1)^k\left(
\begin{array}{c}
m\\
k
\end{array}
\right)B(x_j-x_i^0+k,t)
\label{20}
\end{equation}
because the right-hand side is the result of action of the operator
$(1-\hat{a}_i)^m$.
Note that Eq.(\ref{18}) coincides with Eq.(\ref{3}) and Eq.(\ref{20}) with
Eq.(\ref{4}) where the index $m$ can be called "activity".

To find the generating function of two particles on the ring, we map the
trajectories wrapping the cylinder $\Lambda$ on an infinite plane introducing
coordinates $x+nL$, $n$ integer, for equivalent points. We call trajectories of
two particles compatible if there is at least one possibility to draw them
without intersections. Given starting points $x_1^0$ and  $x_2^0$ at $t=0$,
the pairs of compatible trajectories correspond to end points $x_1,x_2$;
$x_2,x_1+L$;...;$x_1+nL,x_2+nL$; $x_2+nL,x_1+(n+1)L$;... at time $t$. The
index of
activity of compatible trajectories is $0$. A trajectory ending at $x_1+nL$
may interact with the trajectories ending at  $x_2+(n-1)L$ or  $x_2+nL$.
Following the Bethe-ansatz prescription, we should  add to the set of
compatible trajectories two sets of intersecting trajectories taken with minus
sign: the  set which is obtained by permuting end points $x_2+(n-1)L$ and
$x_1+nL$ and the second one obtained by permuting $x_1+nL$ and  $x_2+nL$. Each
new trajectory, in
its turn, interacts with neighbouring trajectories and we should permute their
end points again. Continuing this procedure, we obtain all possible pairs of
trajectories of the first particle with end point $x_1+n_1L$,
the second one with end point $x_2+n_2L$, or vice versa, for arbitrary
integer $n_1$ and $n_2$. The permutations of end points
can be expressed by the determinant as can be seen in Eq.(\ref{6}). The sign
of a pair is defined
by the number of permutations needed to obtain given pair from an compatible
one. The index of activity of each trajectory is defined by the number of
overtakes to the moment $t$.
Evaluation of the number of permutations gives the pre-factor in Eq.(\ref{6}).
The number of overtakes is given by Eq.(\ref{8}).

If the number of particles $P \geq 3$, the elementary squares shown in Fig.1
may occur several times in one horizontal strip of $\Lambda$. If  squares
filled by interacting trajectories are separated one from another by a gap of
empty sites, the above arguments can be applied to each pair of interacting
trajectories separately. The crucial case for the Bethe ansatz is a situation
when the elementary squares are nearest neighbors. The specific property of
the totally ASEP is that, in each pair of interacting trajectories, the right
trajectory remains free and interacts with the next trajectory independently
on its left neighbors. Therefore, we can analyse the interaction between
particles considering successively elementary squares in each row from left to
right  starting from an arbitrary empty square and
then from the top to bottom of the lattice until all unwanted trajectories on
$\Lambda$ will
be removed. As above, all trajectories ending at points  $x_i+n_iL ,i=1,...,P$
must be involved in consideration. Then, Eq.(\ref{6}) is a straightforward
generalisation of the case $P=2$, and Eq.(\ref{8})gives the index of activity
for an arbitrary number of intersecting trajectories. A new element
in the many-particle case is that each trajectory may get $m$ active
intersections and $n$ passive ones.The operator form of Eq.(\ref{19}) and
Eq.(\ref{20}) shows that the resulting activity in this case is $m-n$.

Because of conditions $B(N,t)=0$ for $N<0$, the infinite summations
in Eq.(\ref{3}) and Eq.(\ref{6}) are actually finite for finite $t$.
An advantage of the discrete formulation of the ASEP is a possibility to
illustrate each step of derivation by simple examples. For instance, in the
case $P=2,L=3,t=6,x_1^0=x_1=1,x_2^0=x_2=2$, the generating function is
$G_2=z^{12}+20z^{9}y^{2}+30z^{9}+z^{6}y^{6}+30z^{6}y^{5}+90z^{6}y^{4}
+20z^{6}y^{3}+20z^{3}y^{6}+30z^{3}y^{5}+y^6$. All 243 allowed configurations
enter this expression with proper weights. In the general case, the obtained
expressions open a prospect for detailed investigations of non-stationary
solutions for various initial conditions.

The author is grateful to MPI-PKS, Dresden for hospitality.
This work was supported in part by SNSF Grant No.7SUPJ62295.

*~~ Electronic address: priezzvb@thsun1.jinr.ru


\begin{references}
\bibitem{Liggett}T.M.Liggett,{\it Interacting Particle
Systems} (Springer Verlag,  New York, 1985).
\bibitem{Spohn} H.Spohn,{\it Large Scale Dynamics of Interacting Particles}
(Springer-Verlag, New York, 1991).
\bibitem{Krug} J.Krug, Phys.Rev.Lett. {\bf 67},1882 (1991).
\bibitem{Gva}L.H.Gva and H.Spohn, Phys.Rev.Lett. {\bf 68},725 (1992).
\bibitem{DerEvMuk}B.Derrida, M.R.Evans, and D.Mukamel, J.Phys. A {\bf 26}4911
(1993).
\bibitem{JanLeb}S.A.Janovsky and J.L.Lebowitz, Phys.Rev.A {\bf 45},618 (1992).
\bibitem{Derrida}B.Derrida, Phys.Rep. {\bf 301}, 65 (1998).
\bibitem{Dhar}D.Dhar, Phase Trans. {\bf 9}, 51 (1987).
\bibitem{DerMal}B.Derrida,K.Mallick, J.Phys. A {\bf 30}1031 (1993).
\bibitem{Schutz}G.M.Sch\"utz, J.Stat.Phys. {\bf 88},427 (1997).
\bibitem{Baxter}R.J.Baxter {\it Exactly Solved Models in Statistical
Mechanics}( Academic, New York, 1982).
\bibitem{Korepin}V.E.Korepin, Comm.Math.Phys. {\bf 86},391 (1982).
\bibitem{Izergin}A.G.Izergin, Sov.Phys.Dokl.{\bf 32},878(1987).
\end{references}
\end{document}